\documentclass[aps,prb,twocolumn,superscriptaddress,showpacs,floatfix]{revtex4}

\usepackage{graphicx,color}
\usepackage{amsfonts}
\usepackage[figuresright]{rotating}  
\usepackage{amssymb}
\usepackage{amsmath}
\usepackage{psfrag}
\usepackage{subfigure}
\usepackage{multirow}
\usepackage{tabularx}
\usepackage{textcomp}
\usepackage{units}

\def\nn{\nonumber}

\def\beq{\begin{eqnarray}}
\def\eeq{\end{eqnarray}}
\def\c{\hspace{2pt}}

\renewcommand{\v}[1]{\ensuremath{\mathbf{#1}}} 
 
\let\baraccent=\= 
\renewcommand{\=}[1]{\stackrel{#1}{=}} 

\begin{document}


\title{Wigner Crystallization in Two Dimensions:\\
 Evolution from Long- to Short-Ranged Forces. }

\author{Benjamin M.\ \surname{Fregoso}}
\affiliation{Joint Quantum Institute and Condensed Matter Theory
  Center, Department of Physics, University of Maryland, College Park,
  Maryland 20742-4111, USA} 
\author{C. A. R. \ \surname{S\'{a} de Melo}}
\affiliation{School of Physics, Georgia Institute of Technology, Atlanta, 
Georgia 30332, USA}

\begin{abstract}
\text{}
We study fermions in two dimensions interacting via a long-ranged $1/r$ potential  
for small particle separations and a short-ranged $1/r^3$ potential 
for larger separations in comparison to a length scale $\xi$. We compute the energy of the Wigner crystal 
and of the homogeneous Fermi liquid phases using a variational approach, and determined the 
phase diagram as a function of density and $\xi$ at zero temperature.
We discuss the collective modes in the Fermi liquid phase,  
finite temperature effects on the phase diagram, and possible experimental realizations of this model.
\end{abstract}

\pacs{71.10.Ca,71.10.Pm, 05.30.Fk}
\maketitle

\section{Introduction}
Inhomogeneity plays an important role in many highly correlated 
materials\cite{Wigner1934,Grimes1979,Yoon1999,Kivelson2003,Tanatar1989,Drummond2009,Spivak2003,Spivak2004}. 
For example stripe phases which break translational symmetry of real space in one 
direction have been observed in high temperature superconductors\cite{Kivelson2003}.
Similarly, a Wigner crystal (WC) phase\cite{Wigner1934}   
which has only discrete translational symmetry of triangular 
lattices and six-fold rotational symmetry  
has been observed in electrons on the surface of liquid helium\cite{Grimes1979,Rousseau2009}
and in ultra-clean two-dimensional hole gases\cite{Yoon1999}.

For interactions which decay as the power law  $1/r^\alpha$, a classical argument 
shows that the potential energy scales
with density as $n^{\alpha/2}$ in two dimensions (2D) and the kinetic energy 
as $n$. The ratio of the potential to the kinetic energy scales as
$n^{\alpha/2-1}$ which becomes $n^{-1/2}$,  for Coulomb interactions, a constant
for $1/r^2$ and $n^{1/2}$ for $1/r^3$ potentials.
A stable WC crystal phase occurs when the interaction energy is greater than the 
kinetic energy\cite{Wigner1934}. Hence, a WC is expected in the high density regime or low density regime 
depending on the power law decay of interactions. For Coulomb forces  
this suggests that there is a phase transition at decreasing densities from  
a homogeneous Fermi liquid (FL) phase, which is conducting, 
to a WC phase, which is insulating. In this regime, perturbative methods
fail and one must resort to numerical studies~\cite{Tanatar1989,Drummond2009}.

In reality, many important systems do not have pure power law interactions.
For two-dimensional electron gases (2DEGs) in semiconductor inversion layers 
or quantum wells, the presence of a nearby gate modifies the Coulomb interaction.
Image charges emerge at the gate and screen the Coulomb ($1/r$) interaction, 
which become $1/r^3$ at large distances beyond a length scale given by the distance to the gate.
The effects of such long-ranged forces changes 
the energetics of the 2DEG\cite{Skinner2010}. Recently, the effects of long-ranged 
forces on superconductivity have been explored too\cite{Raghu}.
A similar experimental scenario occurs in ionic liquid transistors\cite{Cho2008,Loth2010},
where an electrolyte is used as a dielectric in a standard field effect 
transistor configuration. Positive and negative charges accumulate 
at opposite ends of the electrolyte. Finally, recent progress in cooling 
techniques have allowed the study of degenerate 
dipolar gases\cite{Lu2012,Ni2008,Baranov2012,Maeda2013,Sun2010,Lin2010,Matveeva2012,Babadi2012,Babadi2011} 
including Wigner crystallization of dipoles with $1/r^3$ potentials\cite{Matveeva2012} and collective 
modes\cite{Babadi2012,Babadi2011,Lu2012a,Kestner2010}. Such objects interact as $1/r^3$ at large separations. 
At short distances the interaction potential is modified by the interactions between the 
electronic clouds.
\begin{figure}
\subfigure{\includegraphics[width=0.45\textwidth]{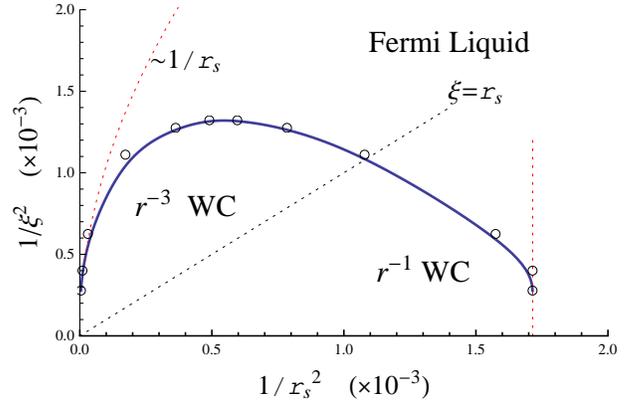}}
\caption{(Color online) Phase diagram of fermions with two-body interaction Eq.~(\ref{eqn:two-body-int}). 
The phase boundary between Wigner crystal (WC) and uniform Fermi liquid (FL) phase as a function of
the screening length $\xi$ in units of $a_0$ and density $n=1/(\pi a_0^2 r_s^2)$. Circles represent
numerical points with 55 electrons and the solid line is a guide to the eye.  
Also shown is the $\xi = r_s$ line, which divides the Coulomb $1/r$ 
regime (right) from the $1/r^3$ regime (left).}
\label{fig:zero-temperature-phase-diagram}
\end{figure}
The common feature of these systems is that there is a length scale in 
the interactions. They are composed of charges of opposite signs which are 
located in layers separated by a dielectric. As such, the interparticle potential crosses over  
between two different power law regimes. 

In this work, we study the ground state properties and collective modes of  
polarized electrons in 2D with screened Coulomb interactions. 
Such system is realized in 2DEG in the presence of 
a nearby gate and a magnetic field parallel to the surface to avoid 
significant orbital effects. We also describe the 
related problem of a gas of fermionic dipoles which have a finite 
size. To be concrete, we consider an interaction 
interpolating smoothly between a 
$1/r$ ($1/r^3$) potential at short (long) distances 
\begin{equation}
\label{eqn:two-body-int}
V(r)  = \frac{ e^2}{\epsilon r} - \frac{ e^2}{\epsilon (r^2+\xi^2)^{1/2}},
\end{equation}
where $\xi$ represents the screening length. While three dimensional effects have 
been investigated in a variety of contexts\cite{Rastelli2006}, here, we consider a 2D model.
Our phase diagram agrees with studies of 2DEGs with a nearby gate\cite{Spivak2003, Spivak2004}, 
and of electrons on the surface of liquid helium\cite{Peeters1983}.
In our analysis, we consider 
a many-body wave function in the form of a Slater determinants and variational 
single-particle wave functions
for the FL and WC phases. In the WC phase, the variational parameter 
is the spatial extent of the single-particle wave functions. Optimizing 
the total energy at zero temperature we explicitly calculate and compare 
the ground state energies of the WC and FL phases.
We note that the effects of screened interactions on the capacitance of 2DEGs and ionic liquids 
have also been explored\cite{Skinner2010,Loth2010}.

The corresponding zero-temperature $(T = 0)$ phase diagram 
is presented in Fig.~\ref{fig:zero-temperature-phase-diagram} as a function of the 
length scale $\xi$ and density $n$ parametrized by $r_s=1/(a_0\sqrt{\pi n})$,
where $a_0=\epsilon \hbar^2/(m e^2)$ is the Bohr radius in the presence of 
the dielectric constant $\epsilon$. The FL and WC phases are indicated in the phase
diagram and are labeled according to the dominant interaction regimes.
For fixed $\xi> \xi_c$ there is a regime of densities where the Wigner crystal is 
energetically more favorable than a uniform FL. 
We also show the asymptotic forms of the phase boundaries in dashed (red) lines.
In the regime dominated by $1/r^3$, the asymptotic phase boundary corresponds to 
$\xi \sim r_s^{1/2}$, while in the Coulomb case it is $r_s =$ constant $(\approx ~ 24.2)$.
Intuitively, for $1/r^3$ potentials (left of the $\xi = r_s$ line) the kinetic ($ 1/r_s^2$) and 
potential $(e^2 \xi^2/r_s^3)$ energies are comparable when $\xi \sim r_s^{1/2}$, whereas for the $1/r$ regime,
the potential energy scales as $e^2/r_s$ and these energies are comparable when $1/r_s^2$ 
is a constant ($\approx ~ 1.72 \times 10^{-3}$). 
Notice that for $\xi< \xi_c = 27.5 a_0$ there is no WC phase.
The existence of the maximum is a general feature
that also follows from the physical argument described above; if $\xi$
is too small the potential energy cannot be of the same order as the 
kinetic energy. The limit $\xi\to \infty$ (Coulomb regime) is not strictly 
accessible within our method. However, by
extrapolating, we find that the WC phase more favorable  
than the WC phase for $r_s > 24.2$. 

For $\xi< \xi_c$ and increasing density from zero to finite 
values, we find a first order phase transition from 
a dipolar FL to a WC. This phase transition has been studied numerically in the 
context of dipole gases with pure $1/r^3$ forces\cite{Matveeva2012}. With additional increases 
of the density beyond the $\xi = r_s$ line, the 
WC crosses over to a Coulomb WC.  Upon further increase of the density, 
the WC melts into a FL. This reentrant behavior for the 
FL phase is a remarkable characteristic of any 2D  
system interacting via an effective potential which crosses over
from long to short range. Fig.~\ref{fig:zero-temperature-phase-diagram} 
constitutes one of our main results. 
\begin{figure}[]
\subfigure{\includegraphics[width=0.45\textwidth]{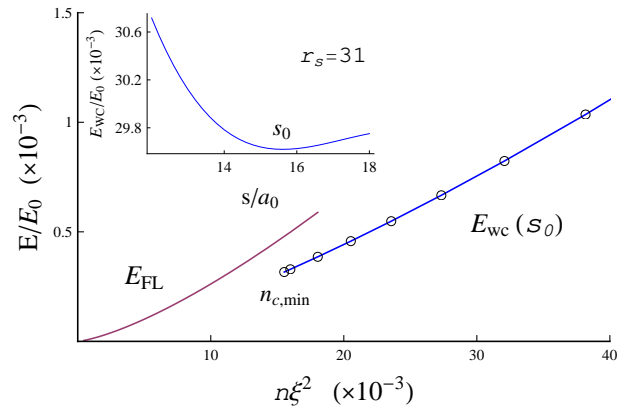}}
\caption{(Color online) Energy per particle vs density (in units of particles per $\xi^2$)
of the WC and FL phases, for $\xi=40 a_0$. See also Fig.~\ref{fig:Ewcvsn}. 
Inset: E$_{\textrm{wc}}(s)$ vs size of the wave function.}
\label{fig:EFL_ESF_E_wc}
\end{figure}

In what follows we show how this phase diagram is obtained. 
In section \ref{sec:model}, we present the Hamiltonian and in section \ref{sec:FL}, 
we compute the energy of the homogeneous FL and discuss its collective modes,
as they are modified by the presence of $\xi$. 
In section \ref{sec:WC}, we obtain the energy in the WC phase with a variational wave function.
In section \ref{sec:disscusion}, we discuss in detail the physical significance of our results, 
and we compare the energies of these two states in different 
density regimes in order to determine the ground state of the system. In addition, we discuss the scaling of 
energy of the WC with respect to density,  give a schematic finite temperature  
phase diagram, and we present our concluding remarks.

\section{Model}
\label{sec:model}
We start from the microscopic Hamiltonian  
\begin{equation}
\label{eqn:Hamiltonian}
\hat{H}=  -\sum_i \frac{\nabla_i^2}{2m}  + \sum_{i<j}{V(\v{r}_i - \v{r}_j)}, 
\end{equation}
where the first term corresponds to the kinetic energy and the second 
term is given in Eq.~(\ref{eqn:two-body-int}), which 
describes the interactions between spinless fermions. The variable 
$\xi$ is the `screening' length and we treat it as a phenomenological parameter.
It could represent a number of physical quantities, e.g., 
twice the distance between a polarized 2DEG and a nearby gate or the 
classical effective size of the dipoles.
Here, $\epsilon$ is the dielectric constant of the medium. Note that for a 
fixed density and in the limit $\xi \to \infty$, the interaction becomes 
$V(r)=e^2/\epsilon r$. Whereas, in the limit $\xi \to 0$, it becomes 
$V(r) = (e\xi)^2/\epsilon r^3$ to lowest non-vanishing order in $\xi$. 
In Section \ref{sec:disscusion}, we also discuss polarized fermionic dipoles 
where there is a factor of two in the two-body interaction.

For both FL and WC phases, we consider the many-body ground state wave function to be 
\begin{align}
\Psi(\v{r}_1,\v{r}_2,...,\v{r}_N) = 
\mathcal{A}\left[\phi_{1}(\v{r}_1)\cdots \phi_{N}(\v{r}_N)\right],
\label{eqn:many-body-wave function}
\end{align}
%
%
where $N$ is the number of particles, and $\mathcal{A}$ is an operator
that describes the anti-symmetrization necessary for indistinguishable fermions. 
Results for indistinguishable bosons and for classical distinguishable 
particles are straight-forward extensions. For simplicity, we assume 
that the spin degrees of freedom are frozen, 
such that the wave function describing individual particles
$\phi_{\ell}({\bf r}_\ell)$ are either plane waves in the FL phase,
or Gaussian centered at individual sites of a triangular lattice
in the Wigner crystal phase. 

\section{Fermi liquid phase}
\label{sec:FL} 
To begin, let us consider the homogeneous FL phase. 
Our system has two length scales. The first scale is $\xi$, 
while the second is the interparticle distance $k_F^{-1}$ set by the density
$n = k_F^2/(4\pi)$, where $k_F$ is the Fermi momentum of 
the FL with Fermi energy $E_F = k_F^2/2m$. 
The relative size of these length scales determines the range of the
potential. Using the relation between $n$, $k_F$ and $r_s$, we see 
that for $\xi/(a_0 r_s) \ll 1$ the fermions interact with a 
$1/r^3$ potential, whereas for $\xi/(a_0 r_s) \gg 1$ 
they interact with the Coulomb $1/r$ potential (see Fig.~\ref{fig:zero-temperature-phase-diagram}).
The energy per particle of a uniform and polarized FL is 
\begin{eqnarray}
\frac{E_{FL}}{E_0} &=& \frac{2}{r_s^2} + \frac{2\bar{\xi}}{r_s^2}  
- \frac{1}{\bar \xi} 
\bigg[
I_{0}(4 \bar{\xi} /r_s )
- I_{2}(4 \bar{\xi}/ r_s) \nonumber \\
&&
+
\frac{8\bar{\xi}}{\pi r_s}  - 1
- L_{0}(4 \bar{\xi}/ r_s) 
+ L_{2}(4 \bar{\xi}/ r_s)
\bigg],
\label{eqn:fermi-liquid-energy}
\end{eqnarray}
%
%
where ${\bar \xi} = \xi/a_0$ is scaled by the Bohr radius $a_0=\epsilon\hbar^2/me^2$,
and the energy $E_{FL}$ is scaled by the Bohr energy 
$E_0 = e^2/2a_0$. The functions $I_n (x)$ are the $n^{th}$ order modified Bessel function 
of the 1st kind and $L_n (x)$ are the $n^{th}$ order Struve functions.
In obtaining Eq.~(\ref{eqn:fermi-liquid-energy}), we used the fact that 
the two-body interactions shown in Eq.~(\ref{eqn:two-body-int}) 
have Fourier transform $V(q) = (2\pi e^2/\epsilon q)(1- e^{-\xi q})$, where $q$
is the magnitude of the two dimensional momentum vector.

The first term in Eq.~(\ref{eqn:fermi-liquid-energy}) is the kinetic energy and 
scales as the density $n$. The second term in
Eq.~(\ref{eqn:fermi-liquid-energy}) is the Hartree contribution,
which also scales as $n$. In the limit 
$\bar{\xi} \to \infty$, $E_{FL}/E_0 = 2/r_s^2 + 2\bar{\xi}/r_s^2 - 16/(3\pi r_s)$ 
is formally divergent as interaction potential is long-ranged, that is, it behaves as $1/r$.
In a 2DEG the positive background charge cancels the Hartree term
and, in this case, $E_{FL}/E_0 = 2/r_s^2 - 16/(3\pi r_s)$, 
in agreement with known results\cite{Rajagopal1977}. 
The last term within the square brackets
is the Fock contribution, which becomes $-2\bar{\xi}/r_s^2 + 256\bar{\xi}^2/(45\pi r_s^3)$
as $\bar{\xi} \rightarrow 0$ and leads to  $E_{FL}/E_0 = 2/r_s^2 + 256\bar{\xi}^2/(45 \pi r_s^3)+ O(\xi^3)$
in this $1/r^3$ regime. To $O(\xi)$ the Fock term cancels 
the Hartree term for pure $1/r^3$ potentials. Physically, the system 
behaves as a charge neutral FL of dipoles. We note that including 
a background to enforce charge neutrality, as in 
2DEGs with $1/r$ potentials, leads to negative energies 
and a self-bound system. However, in the present case the energies are 
positive and the system has positive pressure. The energy of 
the uniform FL is shown in Fig.~\ref{fig:EFL_ESF_E_wc} for 
$\xi=40 a_0$. 

\subsection{Collective modes}
The length scale $\xi$ in the potential introduces important 
modifications to the collective excitations in the FL phase. 
The Fourier transform of Eq.~(\ref{eqn:two-body-int}) at zero 
momentum is well-defined,  $V(0)=2\pi e^2\xi$. 
For these kinds of forces, it is very important to keep the Hartree and Fock 
terms in any approximation to the self-energy\cite{Fregoso2010} as this provides 
a conserving approximation\cite{Kadanoff1962}. 
The collective modes are given by the zeros of the dielectric function
$\epsilon(q,\omega)= 1- V(q)\Pi(q,\omega)$. A non-conserving 
RPA calculation gives,
%
%
\beq
\omega_q = v_F (\bar{\xi}/2)^{1/2}~q,
\eeq
for $\bar{\xi} \gg 1$ and $\omega_q =  v_F[1+ \bar{\xi}^2/2]~q$
for $\bar{\xi}\ll 1$. Hence, we expect the system to support zero sound modes. 
A detailed calculation likely renormalizes the zero sound velocity, but does not alter 
the qualitative physical phenomenon; see, for example, the analysis  
performed in the context of polarized dipole gases\cite{Babadi2012,Babadi2011,Lu2012a,Kestner2010}.
The speed of the zero sound diverges when $\xi \to \infty$, since the power expansion in 
$q/\omega_q$ breaks down in this limit. Indeed, the 
plasmon dispersion relation in 2DEGs has been extensively studied in 
GaAs\cite{Eriksson2000,Hwang2007} semiconductor quantum wells. 
The Fourier transform of the pure Coulomb potential 
$V(q)\sim 1/q$ diverges at zero momentum and leads to 
gapless $\sim\sqrt{q}$ mode dispersion. 

According to the variational principle, the energy obtained using  
the many-body wave function described in Eq.~(\ref{eqn:many-body-wave function}) 
is a rigorous upper bound for the true ground state energy of the system. 
In the following analysis, we perform a variational calculation of 
the ground state energy of the WC phase and compare it with
that of the FL phase obtained in Eq.~(\ref{eqn:fermi-liquid-energy}).

\section{Wigner crystal phase}
\label{sec:WC}
In the Wigner crystal phase, the single-particle wave 
functions are localized at sites $i$ of a 2D triangular 
lattice, $\phi_i (\v{r}) = [1/(s \sqrt{\pi})]\exp\left[-(\v{r}-\v{R}_i)^2/(2 s^2)\right],$
where ${\bf R}_i$ is the site position and $s$ parametrizes the 2D size of the
wave function. These single-particle wave functions are
approximately orthonormal, since the overlap at different 
sites is exponentially small, 
$
\int d^2 r \c \phi^{*}_i(\v{r}) \phi_j(\v{r}) 
= 
\exp
\left[
-(\v{R}_i - \v{R}_j)^2/4 s^2
\right] 
=
\exp
\left[
-R_{ij}^2/(4 s^2)
\right], 
$
provided that the separation $R_{ij}$ between sites $i$ and $j$
is much larger than the extent $s$ of the single-particle wave function.
We denote the lattice spacing by $l$ and consider 
the regime of weakly overlapping single-particle wave functions, 
where $ s < l/2$. The localized nature of the many-body wave function 
is reasonable only in this regime. Explicitly, the energy 
per particle in the WC phase is,
\begin{eqnarray}
\frac{E_{\text{wc}}(\bar{s})}{E_0} & = & 
\frac{1}{\bar{s}^2} 
+ \int_{0}^{\infty} d\bar{k}~ (1- e^{- \bar{\xi} \bar{k}}) e^{-\bar{k}^2 \bar{s}^2 /2} F_1(\bar{k}) 
\nonumber \\
&& - \frac{\sqrt{2\pi}}{2 \bar{s}} 
\left( 1- e^{\bar{\xi}^2/2\bar{s}^2}\mathrm{erfc}(\frac{\bar{\xi}}{\sqrt{2}\bar{s}}) \right) 
F_2(\bar{s}), 
\label{eqn:wigner-crystal-energy}
\end{eqnarray}
where the functions $F_i (y)$ are lattice sums given by 
$F_1 (\bar{k}) = \sum_{i\neq j} J_0(\bar{k} \bar{R}_{ij})/N$ appearing in 
the second term of Eq.~(\ref{eqn:wigner-crystal-energy}),
and $F_2 (\bar{s}) = \sum_{i\neq j}e^{-\bar{R}_{ij}^2/2\bar{s}^2}/N$
appearing in the third term of Eq.~(\ref{eqn:wigner-crystal-energy}). Here, we defined the dimensionless variables
$\bar{s}=s/a_0$, $\bar{k}=k a_0$, $\bar{R}_{ij}=R_{ij}/a_0$ and $J_{n}$ is the $n^{th}$ 
order Bessel function of the first kind. The first term is the zero point motion due to the localization 
of the particles at the lattice sites. This term favors large 
single-particle wave functions. The second and third terms are the Hartree-Fock (HF) 
contribution (respectively). The HF contribution is non-monotonic in $s$. The extension 
to particles with bosonic statistics results in a change of the sign of the Fock 
term whereas for classical particles only the Hartree term is present.

We comment on the limiting behavior of Eq.~(\ref{eqn:wigner-crystal-energy}).
Fermions interact via a Coulomb potential when $\xi \to \infty$. 
In this limit, the ground state energy is
\beq
\frac{E^{\text{Coulomb}}_{\text{wc}}}{E_0} &=& \frac{1}{\bar{s}^2} + \frac{\sqrt{2\pi}}{2 \bar{s} N} \sum_{i\neq j} e^{-\bar{R}_{ij}^2/4\bar{s}^2} I_{0}(\bar{R}_{ij}^2/4\bar{s}^2)\nn \\
&&-  \frac{\sqrt{2\pi}}{2 \bar{s} N} \sum_{i\neq j} e^{-\bar{R}_{ij}^2/2\bar{s}^2}.
\label{eqn:ColoumbWCenergy}
\eeq
In the regime of non-overlapping wave functions $l/2s \gg 1$, the 
asymptotic form of the Bessel function is $I_0(x)\sim e^{x}/\sqrt{2\pi x}$.  
As we see, the series in the second (Hartree) term  diverges when $N\to\infty$
as expected, since we have not included a neutralizing background charge.
In this case, taking a finite $N$ gives a well defined energy which 
is non-monotonic in $s$.  We find that this behavior extends 
to the regime of $\xi < \infty$ and produces a WC state 
energy $E_{\rm wc}(s)$ that does not have a minimum as a function 
of $s$, when the density is either too small or too large.
On the other hand, in the limit of $\xi\to 0$, we 
obtain that the $O(\xi)$ contribution from the Fock  
term cancels the Hartree term (as in the FL phase) and the first non-vanishing 
term is $O(\xi^2)$. Since the resulting expression is not very illuminating, we 
omit it here. 

\begin{figure}
\subfigure{\includegraphics[width=0.45\textwidth]{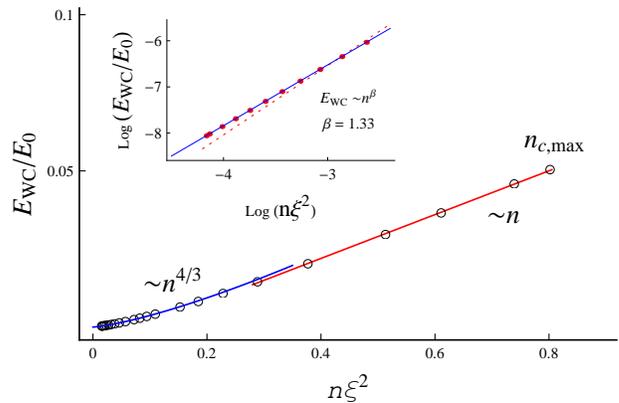}}
\caption{(Color online) Scaling of the energy $E_{\textrm{wc}}$ of the Wigner crystal (WC) near $n_{c,min}$ and $n_{c,max}$. 
Here, $\xi= 40 a_0$. The picture shows two different scalings with density. The crossover 
occurs at $r_s=\xi/a_0$ where $n\xi^2=1/\pi$. Inset shows a fit 
at low densities (blue line) and the classical $\sim n^{3/2}$ scaling of the WC (red-dashed line).}
\label{fig:Ewcvsn}
\end{figure}
\section{Discussion and conclusions}
\label{sec:disscusion}
For $\xi < \infty$, the integral appearing in Eq.~(\ref{eqn:wigner-crystal-energy}) 
needs to be calculated numerically. We computed the energy 
per particle as a function of density $n = 2/(l^2 \sqrt{3})$ in the WC
phase for a 2D triangular array of $43$ and $55$ particles. 
In Fig.~\ref{fig:zero-temperature-phase-diagram}, 
we show the resulting phase diagram for 55 particles, 
but we would like to emphasize that the qualitative behavior
found is essentially unchanged for higher number of particles.
The calculation of phase boundaries for the full range
of parameters $\xi$ and $r_s$ is very intensive 
when it involves a large number of particles, however, we do not 
expect any major qualitative changes in behavior.

We establish the phase boundary between solid and liquid
phases by minimizing the energy in the solid phase with respect to 
the variational parameter $s$. A typical behavior of the energy 
as a function $s$ is shown in Fig.~\ref{fig:EFL_ESF_E_wc}.
When there is a local minimum of $E_{\rm wc} (s)$ located  
at $s = s_0$, we compare the energy $E_{\rm wc} (s_0)$ with the 
energy of the screened FL with the same density. 
We find that the energy of the WC is always lower than the FL energy 
in the regime of tested values of $\xi$. However, a local minimum of 
$E_{\rm wc} (s)$ does not exist for densities below a minimum 
$n_{c,min} (\xi)$ and for densities above a maximum 
$n_{c,max} (\xi)$, where the WC phase is unstable. 
This establishes that the WC phase is unstable for densities
$n$ satisfying the condition $n < n_{c,min} (\xi)$ and
$n > n_{c,max} (\xi)$, in which case, the FL phase is the stable phase. 

%
\begin{figure}
\subfigure{\includegraphics[width=0.45\textwidth]{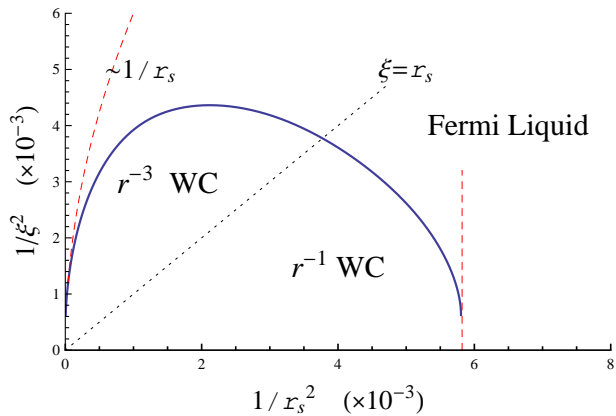}}
\caption{(Color online) Phase diagram of dipoles with  
effective size $\xi$ in 2D, versus density ($1/r_s^2$). 
The Fermi liquid (FL) and Wigner Crystal (WC) phases are indicated.}
\label{fig:zero-temperature-phase-diagram2}
\end{figure}
A phase transition from a ferromagnetic FL to a ferromagnetic WC 
phase in a 2DEG with $1/r$ interactions is expected to occur 
at $r_s = 29$ according to 
recent Monte-Carlo simulations\cite{Drummond2009}. 
In the $\xi \to \infty$ limit, we find such a phase 
transition at $r_s =24.2$, with finite-sized samples. 
We note that our results do not rigorously apply in this regime since  
interactions are of pure Coulomb character and a neutralizing 
background charge must be explicitly considered. 
Here, we are interested in the regime of $\xi < \infty$,
where our results are expected to be qualitatively correct.

We note that the optimized energy values in the WC can be fitted 
with a power law 
\beq
E_{\textrm{wc}}(s_0) \sim   n^{4/3}
\eeq
in the low density regime, i.e., $r_s \gg \xi$ where interactions are of $1/r^3$ character. 
This is shown explicitly in Fig.~\ref{fig:Ewcvsn} for $\xi = 40 a_0$. We 
tested the scaling for $40 < \xi/a_0 < 60$ to within $1\%$ error, but we 
cannot completely rule out a crossover behavior\cite{Lieb2005}. 
A similar $n^{4/3}$ scaling behavior has been studied in 2DEGs with a 
$1/r^3$ potential and spin-orbit interactions\cite{Berg2012}. 
In the high density regime ($r_s\ll \xi$) the energy is a linear power law
of the density $E_{\textrm{wc}} = A \vert n - n_{c, max}\vert + B $,
where interaction are of $1/r$ character. 
This scaling persists in the same tested range and within the same error. 

As mentioned earlier, the potential shown in Equation Eq.~(\ref{eqn:two-body-int}) 
describes the interactions between polarized 
electrons in a clean 2DEG with a nearby gate. As is well known, 
by applying a magnetic field parallel to the surface one avoids significant orbital 
effects and a one component 2DEG is obtained.
However, disorder is always present and makes it difficult to reach the low 
density regime where the reentrant FL is predicted.

In contrast, another system where disorder does not play a role, 
corresponds to degenerate polarized dipolar Fermi gases\cite{Lu2012,Ni2008}.
We extended our results to describe a system of dipoles in 2D 
whose centers move in the plane perpendicular
to the polarization axis\cite{Mitra}. In this case, 
$\xi$ could parametrize a hard core radius (or size of the molecule) 
below which interactions are no longer of the $1/r^3$. 
In this new situation, the interaction potential of Eq.~(\ref{eqn:two-body-int}) 
acquires a prefactor of two, but the calculations are entirely analogous.
In Fig.~\ref{fig:zero-temperature-phase-diagram2} 
we show the phase diagram for 43 dipoles. We observe the same features
as for screened 2DEGs. For $\xi< \xi_c = 15 a_0$ there is no
stable WC phase. In the Coulomb limit the WC is stable for $r_s > 13 a_0$.
In the expressions for the energy of the FL and WC a prefactor of 2 is 
needed in the interaction energy. The collective modes dispersion is 
$\omega_q=v_F \bar{\xi}^{1/2} q$ for $\bar{\xi}\gg 1$ and 
$\omega_q= v_F [1+ 2 \bar{\xi}^2] q$ for $\bar{\xi}\ll 1$. 
Our results suggest that realistic dipole gases with size $\sim 10 a_0$, 
are Fermi liquids at low \textit{and} very high densities. 
We provided estimates for the critical 
densities as a function of $\xi$ in a simplified model. 
If only $1/r^3$ interactions are considered\cite{Matveeva2012} 
one expects a WC instability with increasing densities near the dashed red 
line on the left in Fig.~\ref{fig:zero-temperature-phase-diagram2}, where 
the interaction energy is of the order of the kinetic energy. However this 
red dashed line is strongly modified (blue line), when the 
hard core is considered.

\begin{figure}
\subfigure{\includegraphics[width=0.45\textwidth]{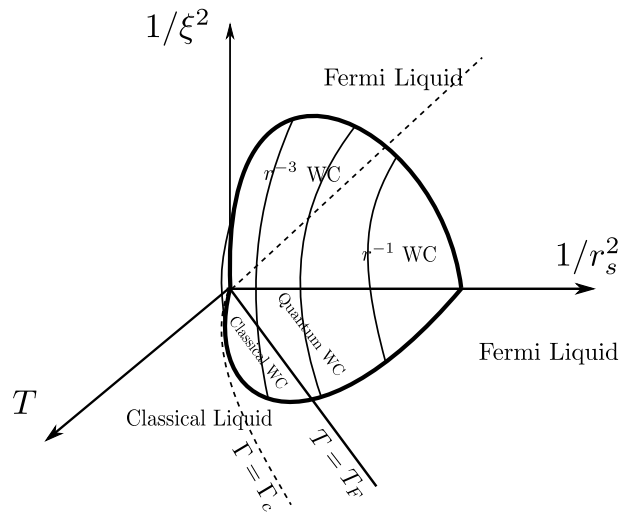}}
\caption{Schematic phase diagram as function of 
density $\sim 1/r_s^2$, temperature $T$ and screening parameter $\xi$ 
in appropriate units. $\Gamma$ is the ratio of the potential to kinetic energy. See text for details.}
\label{fig:phase_diagram_T}
\end{figure}
\subsection{Finite temperature phase diagram}
\label{sec:finiteT}
Next, we discuss the finite temperature phase diagram 
of the 2DEG with screened interactions. The assumption is that 
the interactions can be modeled with the simple form Eq.~(\ref{eqn:two-body-int}).
At finite temperatures, we expect the melting of the WC 
to a liquid-like state\cite{Nelson1979}.
In Fig.~\ref{fig:phase_diagram_T}, we show the phase diagram 
in the ($T, n, 1/\xi^2$) parameter space. In the $1/\xi^2=0$ plane,
the thermodynamics of the classical 2DEG is determined by the quantity 
$\Gamma$, which is the ratio of the interaction energy to the kinetic energy per particle.
For a classical 2DEG, this ratio is $\Gamma = (\pi n)^{1/2}e^2/T$. 
For $\Gamma <1 $, the kinetic energy dominates and the system behaves as 
a classical liquid. For $\Gamma \gg 1$, the Coulomb interaction dominates and 
we expect a classical WC.  The phase boundary is given by the criterion\cite{Grimes1979}
$\Gamma= \Gamma_c \sim 100$. For temperatures below $E_F\sim n\sim 1/r_s^2$, we obtain 
a quantum WC. At high densities the melting temperature is eventually driven down
with increasing density as the Pauli principle prevents strong correlations. 
There is a quantum phase transition in the regime with pure Coulomb interactions  
at $r_s \approx 24.2$. This transition can be seen along the $1/r_s^2$ line
in Fig.~\ref{fig:phase_diagram_T},  where $\Gamma \sim n^{-1/2}$ in the Hartree-Fock approximation. 
This defines a critical density above which $\Gamma <1$ and 
a quantum fluid is recovered. The full melting line and the quantum/classical 
crossover in the plane $1/\xi^2=0$ is sketched in 
Fig.~\ref{fig:phase_diagram_T}, see also \cite{Hwang2007}. Connecting
the WC phases of the $1/\xi^2=0$ and $T=0$ planes is a dome,  
where a WC is stable. Within this dome there exists at least four 
regions corresponding to quantum vs classical and $1/r$ vs $1/r^3$ regimes. 
Outside the dome there are only quantum and classical liquid-like phases. \\

In conclusion, we have obtained the phase diagram of  
fermions with interactions which interpolate smoothly between 
short-ranged and long-ranged regimes in 2D. The phase diagram obtained 
is generic to any system with such crossover, e.g, clean 2DEGs 
in semiconductor inversion layers or quantum wells in the presence 
of a screening gate, and polarized dipolar gases. 

\begin{acknowledgments} 
We thank V. Yakovenko for helpful disscussions and the JQI-PFC (BMF) and ARO (W911NF-09-1-0220) (CSdM) for support.
\end{acknowledgments} 

\bibliographystyle{apsrev}

\begin{thebibliography}{35}
\expandafter\ifx\csname natexlab\endcsname\relax\def\natexlab#1{#1}\fi
\expandafter\ifx\csname bibnamefont\endcsname\relax
  \def\bibnamefont#1{#1}\fi
\expandafter\ifx\csname bibfnamefont\endcsname\relax
  \def\bibfnamefont#1{#1}\fi
\expandafter\ifx\csname citenamefont\endcsname\relax
  \def\citenamefont#1{#1}\fi
\expandafter\ifx\csname url\endcsname\relax
  \def\url#1{\texttt{#1}}\fi
\expandafter\ifx\csname urlprefix\endcsname\relax\def\urlprefix{URL }\fi
\providecommand{\bibinfo}[2]{#2}
\providecommand{\eprint}[2][]{\url{#2}}

\bibitem[{\citenamefont{Wigner}(1934)}]{Wigner1934}
\bibinfo{author}{\bibfnamefont{E.}~\bibnamefont{Wigner}},
  \bibinfo{journal}{Phys. Rev.} \textbf{\bibinfo{volume}{46}},
  \bibinfo{pages}{1002} (\bibinfo{year}{1934}).

\bibitem[{\citenamefont{Grimes and Adams}(1979)}]{Grimes1979}
\bibinfo{author}{\bibfnamefont{C.~C.} \bibnamefont{Grimes}} \bibnamefont{and}
  \bibinfo{author}{\bibfnamefont{G.}~\bibnamefont{Adams}},
  \bibinfo{journal}{Phys. Rev. Lett.} \textbf{\bibinfo{volume}{42}},
  \bibinfo{pages}{795} (\bibinfo{year}{1979}).

\bibitem[{\citenamefont{Yoon et~al.}(1999)\citenamefont{Yoon, Li, Shahar, Tsui,
  and Shayegan}}]{Yoon1999}
\bibinfo{author}{\bibfnamefont{J.}~\bibnamefont{Yoon}},
  \bibinfo{author}{\bibfnamefont{C.~C.} \bibnamefont{Li}},
  \bibinfo{author}{\bibfnamefont{D.}~\bibnamefont{Shahar}},
  \bibinfo{author}{\bibfnamefont{D.~C.} \bibnamefont{Tsui}}, \bibnamefont{and}
  \bibinfo{author}{\bibfnamefont{M.}~\bibnamefont{Shayegan}},
  \bibinfo{journal}{Phys. Rev. Lett.} \textbf{\bibinfo{volume}{82}},
  \bibinfo{pages}{1744} (\bibinfo{year}{1999}).

\bibitem[{\citenamefont{Kivelson et~al.}(2003)\citenamefont{Kivelson, Bindloss,
  Fradkin, Oganesyan, Tranquada, Kapitulnik, and Howald}}]{Kivelson2003}
\bibinfo{author}{\bibfnamefont{S.~A.} \bibnamefont{Kivelson}},
  \bibinfo{author}{\bibfnamefont{I.~P.} \bibnamefont{Bindloss}},
  \bibinfo{author}{\bibfnamefont{E.}~\bibnamefont{Fradkin}},
  \bibinfo{author}{\bibfnamefont{V.}~\bibnamefont{Oganesyan}},
  \bibinfo{author}{\bibfnamefont{J.~M.} \bibnamefont{Tranquada}},
  \bibinfo{author}{\bibfnamefont{A.}~\bibnamefont{Kapitulnik}},
  \bibnamefont{and} \bibinfo{author}{\bibfnamefont{C.}~\bibnamefont{Howald}},
  \bibinfo{journal}{Rev. Mod. Phys.} \textbf{\bibinfo{volume}{75}},
  \bibinfo{pages}{1201} (\bibinfo{year}{2003}).

\bibitem[{\citenamefont{Tanatar and Ceperley}(1989)}]{Tanatar1989}
\bibinfo{author}{\bibfnamefont{B.}~\bibnamefont{Tanatar}} \bibnamefont{and}
  \bibinfo{author}{\bibfnamefont{D.~M.} \bibnamefont{Ceperley}},
  \bibinfo{journal}{Phys. Rev. B} \textbf{\bibinfo{volume}{39}},
  \bibinfo{pages}{5005} (\bibinfo{year}{1989}).

\bibitem[{\citenamefont{Drummond and Needs}(2009)}]{Drummond2009}
\bibinfo{author}{\bibfnamefont{N.~D.} \bibnamefont{Drummond}} \bibnamefont{and}
  \bibinfo{author}{\bibfnamefont{R.~J.} \bibnamefont{Needs}},
  \bibinfo{journal}{Phys. Rev. Lett.} \textbf{\bibinfo{volume}{102}},
  \bibinfo{pages}{126402} (\bibinfo{year}{2009}).

\bibitem[{\citenamefont{Spivak}(2003)}]{Spivak2003}
\bibinfo{author}{\bibfnamefont{B.}~\bibnamefont{Spivak}},
  \bibinfo{journal}{Phys. Rev. B} \textbf{\bibinfo{volume}{67}},
  \bibinfo{pages}{125205} (\bibinfo{year}{2003}).

\bibitem[{\citenamefont{Spivak and Kivelson}(2004)}]{Spivak2004}
\bibinfo{author}{\bibfnamefont{B.}~\bibnamefont{Spivak}} \bibnamefont{and}
  \bibinfo{author}{\bibfnamefont{S.~A.} \bibnamefont{Kivelson}},
  \bibinfo{journal}{Phys. Rev. B} \textbf{\bibinfo{volume}{70}},
  \bibinfo{pages}{155114} (\bibinfo{year}{2004}).

\bibitem[{\citenamefont{Rousseau et~al.}(2009)\citenamefont{Rousseau, Ponarin,
  Hristakos, Avenel, Varoquaux, and Mukharsky}}]{Rousseau2009}
\bibinfo{author}{\bibfnamefont{E.}~\bibnamefont{Rousseau}},
  \bibinfo{author}{\bibfnamefont{D.}~\bibnamefont{Ponarin}},
  \bibinfo{author}{\bibfnamefont{L.}~\bibnamefont{Hristakos}},
  \bibinfo{author}{\bibfnamefont{O.}~\bibnamefont{Avenel}},
  \bibinfo{author}{\bibfnamefont{E.}~\bibnamefont{Varoquaux}},
  \bibnamefont{and}
  \bibinfo{author}{\bibfnamefont{Y.}~\bibnamefont{Mukharsky}},
  \bibinfo{journal}{Phys. Rev. B} \textbf{\bibinfo{volume}{79}},
  \bibinfo{pages}{045406} (\bibinfo{year}{2009}).

\bibitem[{\citenamefont{Skinner and Fogler}(2010)}]{Skinner2010}
\bibinfo{author}{\bibfnamefont{B.}~\bibnamefont{Skinner}} \bibnamefont{and}
  \bibinfo{author}{\bibfnamefont{M.~M.} \bibnamefont{Fogler}},
  \bibinfo{journal}{Phys. Rev. B} \textbf{\bibinfo{volume}{82}},
  \bibinfo{pages}{201306(R)} (\bibinfo{year}{2010}).

\bibitem[{\citenamefont{Raghu et~al.}()\citenamefont{Raghu, Berg, Chubukov, and
  Kivelson}}]{Raghu}
\bibinfo{author}{\bibfnamefont{S.}~\bibnamefont{Raghu}},
  \bibinfo{author}{\bibfnamefont{E.}~\bibnamefont{Berg}},
  \bibinfo{author}{\bibfnamefont{A.~V.} \bibnamefont{Chubukov}},
  \bibnamefont{and} \bibinfo{author}{\bibfnamefont{S.~A.}
  \bibnamefont{Kivelson}}, \bibinfo{note}{arXiv:1111.2982v1}.

\bibitem[{\citenamefont{Cho et~al.}(2008)\citenamefont{Cho, Lee, Xia, Kim, He,
  Renn, Lodge, and Frisbie}}]{Cho2008}
\bibinfo{author}{\bibfnamefont{J.~H.} \bibnamefont{Cho}},
  \bibinfo{author}{\bibfnamefont{J.}~\bibnamefont{Lee}},
  \bibinfo{author}{\bibfnamefont{Y.}~\bibnamefont{Xia}},
  \bibinfo{author}{\bibfnamefont{B.}~\bibnamefont{Kim}},
  \bibinfo{author}{\bibfnamefont{Y.}~\bibnamefont{He}},
  \bibinfo{author}{\bibfnamefont{M.~J.} \bibnamefont{Renn}},
  \bibinfo{author}{\bibfnamefont{T.~P.} \bibnamefont{Lodge}}, \bibnamefont{and}
  \bibinfo{author}{\bibfnamefont{C.~D.} \bibnamefont{Frisbie}},
  \bibinfo{journal}{Nature Materials} \textbf{\bibinfo{volume}{7}},
  \bibinfo{pages}{900} (\bibinfo{year}{2008}).

\bibitem[{\citenamefont{Loth et~al.}(2010)\citenamefont{Loth, Skinner, and
  Shklovskii}}]{Loth2010}
\bibinfo{author}{\bibfnamefont{M.~S.} \bibnamefont{Loth}},
  \bibinfo{author}{\bibfnamefont{B.}~\bibnamefont{Skinner}}, \bibnamefont{and}
  \bibinfo{author}{\bibfnamefont{B.~I.} \bibnamefont{Shklovskii}},
  \bibinfo{journal}{Phys. Rev. E} \textbf{\bibinfo{volume}{82}},
  \bibinfo{pages}{056102} (\bibinfo{year}{2010}).

\bibitem[{\citenamefont{Lu et~al.}(2012)\citenamefont{Lu, Burdick, and
  Lev}}]{Lu2012}
\bibinfo{author}{\bibfnamefont{M.}~\bibnamefont{Lu}},
  \bibinfo{author}{\bibfnamefont{N.~Q.} \bibnamefont{Burdick}},
  \bibnamefont{and} \bibinfo{author}{\bibfnamefont{B.~L.} \bibnamefont{Lev}},
  \bibinfo{journal}{Phys. Rev. Lett.} \textbf{\bibinfo{volume}{108}},
  \bibinfo{pages}{215301} (\bibinfo{year}{2012}).

\bibitem[{\citenamefont{Ni et~al.}(2008)\citenamefont{Ni, Ospelkaus,
  de~Miranda, Pe{'}er, Neyenhuis, Zirbel, Kotochigova, Julienne, Jin, and
  Ye}}]{Ni2008}
\bibinfo{author}{\bibfnamefont{K.-K.} \bibnamefont{Ni}},
  \bibinfo{author}{\bibfnamefont{S.}~\bibnamefont{Ospelkaus}},
  \bibinfo{author}{\bibfnamefont{M.~H.~G.} \bibnamefont{de~Miranda}},
  \bibinfo{author}{\bibfnamefont{A.}~\bibnamefont{Pe{'}er}},
  \bibinfo{author}{\bibfnamefont{B.}~\bibnamefont{Neyenhuis}},
  \bibinfo{author}{\bibfnamefont{J.~J.} \bibnamefont{Zirbel}},
  \bibinfo{author}{\bibfnamefont{S.}~\bibnamefont{Kotochigova}},
  \bibinfo{author}{\bibfnamefont{P.~S.} \bibnamefont{Julienne}},
  \bibinfo{author}{\bibfnamefont{D.~S.} \bibnamefont{Jin}}, \bibnamefont{and}
  \bibinfo{author}{\bibfnamefont{J.}~\bibnamefont{Ye}},
  \bibinfo{journal}{Science} \textbf{\bibinfo{volume}{322}},
  \bibinfo{pages}{231} (\bibinfo{year}{2008}).

\bibitem[{\citenamefont{Baranov et~al.}(2012)\citenamefont{Baranov, Dalmonte,
  Pupillo, and Zoller}}]{Baranov2012}
\bibinfo{author}{\bibfnamefont{M.~A.} \bibnamefont{Baranov}},
  \bibinfo{author}{\bibfnamefont{M.}~\bibnamefont{Dalmonte}},
  \bibinfo{author}{\bibfnamefont{G.}~\bibnamefont{Pupillo}}, \bibnamefont{and}
  \bibinfo{author}{\bibfnamefont{P.}~\bibnamefont{Zoller}},
  \bibinfo{journal}{Chem. Rev.} \textbf{\bibinfo{volume}{112 (9)}},
  \bibinfo{pages}{50125061} (\bibinfo{year}{2012}).

\bibitem[{\citenamefont{Maeda et~al.}(2013)\citenamefont{Maeda, Hatsuda, and
  Baym}}]{Maeda2013}
\bibinfo{author}{\bibfnamefont{K.}~\bibnamefont{Maeda}},
  \bibinfo{author}{\bibfnamefont{T.}~\bibnamefont{Hatsuda}}, \bibnamefont{and}
  \bibinfo{author}{\bibfnamefont{G.}~\bibnamefont{Baym}},
  \bibinfo{journal}{Phys. Rev. A} \textbf{\bibinfo{volume}{87}},
  \bibinfo{pages}{021604(R)} (\bibinfo{year}{2013}).

\bibitem[{\citenamefont{Sun et~al.}(2010)\citenamefont{Sun, Wu, and
  Das~Sarma}}]{Sun2010}
\bibinfo{author}{\bibfnamefont{K.}~\bibnamefont{Sun}},
  \bibinfo{author}{\bibfnamefont{C.}~\bibnamefont{Wu}}, \bibnamefont{and}
  \bibinfo{author}{\bibfnamefont{S.}~\bibnamefont{Das~Sarma}},
  \bibinfo{journal}{Physical Review B} \textbf{\bibinfo{volume}{82}},
  \bibinfo{pages}{075105} (\bibinfo{year}{2010}).

\bibitem[{\citenamefont{Lin et~al.}(2010)\citenamefont{Lin, Zhao, and
  Liu}}]{Lin2010}
\bibinfo{author}{\bibfnamefont{C.}~\bibnamefont{Lin}},
  \bibinfo{author}{\bibfnamefont{E.}~\bibnamefont{Zhao}}, \bibnamefont{and}
  \bibinfo{author}{\bibfnamefont{W.~V.} \bibnamefont{Liu}},
  \bibinfo{journal}{Phys. Rev. B} \textbf{\bibinfo{volume}{81}},
  \bibinfo{pages}{045115} (\bibinfo{year}{2010}).

\bibitem[{\citenamefont{Matveeva and Giorgini}(2012)}]{Matveeva2012}
\bibinfo{author}{\bibfnamefont{N.}~\bibnamefont{Matveeva}} \bibnamefont{and}
  \bibinfo{author}{\bibfnamefont{S.}~\bibnamefont{Giorgini}},
  \bibinfo{journal}{Phys. Rev. Lett.} \textbf{\bibinfo{volume}{109}},
  \bibinfo{pages}{200401} (\bibinfo{year}{2012}).

\bibitem[{\citenamefont{Babadi and Demler}(2012)}]{Babadi2012}
\bibinfo{author}{\bibfnamefont{M.}~\bibnamefont{Babadi}} \bibnamefont{and}
  \bibinfo{author}{\bibfnamefont{E.}~\bibnamefont{Demler}},
  \bibinfo{journal}{Phys. Rev. A} \textbf{\bibinfo{volume}{86}},
  \bibinfo{pages}{063638} (\bibinfo{year}{2012}).

\bibitem[{\citenamefont{Babadi and Demler}(2011)}]{Babadi2011}
\bibinfo{author}{\bibfnamefont{M.}~\bibnamefont{Babadi}} \bibnamefont{and}
  \bibinfo{author}{\bibfnamefont{E.}~\bibnamefont{Demler}},
  \bibinfo{journal}{Phys. Rev. A} \textbf{\bibinfo{volume}{84}},
  \bibinfo{pages}{033636} (\bibinfo{year}{2011}).

\bibitem[{\citenamefont{Lu and Shlyapnikov}(2012)}]{Lu2012a}
\bibinfo{author}{\bibfnamefont{Z.-K.} \bibnamefont{Lu}} \bibnamefont{and}
  \bibinfo{author}{\bibfnamefont{G.~V.} \bibnamefont{Shlyapnikov}},
  \bibinfo{journal}{Phys. Rev. A} \textbf{\bibinfo{volume}{85}},
  \bibinfo{pages}{023614} (\bibinfo{year}{2012}).

\bibitem[{\citenamefont{Kestner and DasSarma}(2010)}]{Kestner2010}
\bibinfo{author}{\bibfnamefont{J.~P.} \bibnamefont{Kestner}} \bibnamefont{and}
  \bibinfo{author}{\bibfnamefont{S.} \bibnamefont{DasSarma}},
  \bibinfo{journal}{Phys. Rev. A} \textbf{\bibinfo{volume}{82}},
  \bibinfo{pages}{033608} (\bibinfo{year}{2010}).

\bibitem[{\citenamefont{Rastelli et~al.}(2006)\citenamefont{Rastelli,
  Qu\'{e}merais, and Fratini}}]{Rastelli2006}
\bibinfo{author}{\bibfnamefont{G.}~\bibnamefont{Rastelli}},
  \bibinfo{author}{\bibfnamefont{P.}~\bibnamefont{Qu\'{e}merais}},
  \bibnamefont{and} \bibinfo{author}{\bibfnamefont{S.}~\bibnamefont{Fratini}},
  \bibinfo{journal}{Phys. Rev. B} \textbf{\bibinfo{volume}{73}},
  \bibinfo{pages}{155103} (\bibinfo{year}{2006}).

\bibitem[{\citenamefont{Peeters and Platzman}(1983)}]{Peeters1983}
\bibinfo{author}{\bibfnamefont{F.~M.} \bibnamefont{Peeters}} \bibnamefont{and}
  \bibinfo{author}{\bibfnamefont{P.~M.} \bibnamefont{Platzman}},
  \bibinfo{journal}{Phys. Rev. Lett.} \textbf{\bibinfo{volume}{50}},
  \bibinfo{pages}{2021} (\bibinfo{year}{1983}).

\bibitem[{\citenamefont{Rajagopal and Kimball}(1977)}]{Rajagopal1977}
\bibinfo{author}{\bibfnamefont{A.~K.} \bibnamefont{Rajagopal}}
  \bibnamefont{and} \bibinfo{author}{\bibfnamefont{J.}~\bibnamefont{Kimball}},
  \bibinfo{journal}{Phys. Rev. B} \textbf{\bibinfo{volume}{15}},
  \bibinfo{pages}{2019} (\bibinfo{year}{1977}).

\bibitem[{\citenamefont{Fregoso and Fradkin}(2010)}]{Fregoso2010}
\bibinfo{author}{\bibfnamefont{B.~M.} \bibnamefont{Fregoso}} \bibnamefont{and}
  \bibinfo{author}{\bibfnamefont{E.}~\bibnamefont{Fradkin}},
  \bibinfo{journal}{Phys. Rev. B} \textbf{\bibinfo{volume}{81}},
  \bibinfo{pages}{214443} (\bibinfo{year}{2010}).

\bibitem[{\citenamefont{Kadanoff and Baym}(1962)}]{Kadanoff1962}
\bibinfo{author}{\bibfnamefont{L.~P.} \bibnamefont{Kadanoff}} \bibnamefont{and}
  \bibinfo{author}{\bibfnamefont{G.}~\bibnamefont{Baym}},
  \emph{\bibinfo{title}{Quantum Statistical Mechanics}}
  (\bibinfo{publisher}{Benjamin, New York}, \bibinfo{year}{1962}).

\bibitem[{\citenamefont{Eriksson et~al.}(2000)\citenamefont{Eriksson, Pinczuk,
  Dennis, Hirjibehedin, Simon, Pfeiffer, and West}}]{Eriksson2000}
\bibinfo{author}{\bibfnamefont{M.}~\bibnamefont{Eriksson}},
  \bibinfo{author}{\bibfnamefont{A.}~\bibnamefont{Pinczuk}},
  \bibinfo{author}{\bibfnamefont{B.}~\bibnamefont{Dennis}},
  \bibinfo{author}{\bibfnamefont{C.}~\bibnamefont{Hirjibehedin}},
  \bibinfo{author}{\bibfnamefont{S.}~\bibnamefont{Simon}},
  \bibinfo{author}{\bibfnamefont{L.}~\bibnamefont{Pfeiffer}}, \bibnamefont{and}
  \bibinfo{author}{\bibfnamefont{K.}~\bibnamefont{West}},
  \bibinfo{journal}{Physica E} \textbf{\bibinfo{volume}{6}},
  \bibinfo{pages}{165} (\bibinfo{year}{2000}).

\bibitem[{\citenamefont{Hwang and DasSarma}(2007)}]{Hwang2007}
\bibinfo{author}{\bibfnamefont{E.~H.} \bibnamefont{Hwang}} \bibnamefont{and}
  \bibinfo{author}{\bibfnamefont{S.} \bibnamefont{DasSarma}},
  \bibinfo{journal}{Phys. Rev. B} \textbf{\bibinfo{volume}{75}},
  \bibinfo{pages}{205418} (\bibinfo{year}{2007}).

\bibitem[{\citenamefont{Lieb and Seiringer}(2005)}]{Lieb2005}
\bibinfo{author}{\bibfnamefont{E.~H.} \bibnamefont{Lieb}} \bibnamefont{and}
  \bibinfo{author}{\bibfnamefont{R.}~\bibnamefont{Seiringer}},
  \bibinfo{author}{\bibfnamefont{J.~P.}~\bibnamefont{Solovej}},
  \bibinfo{journal}{Phys. Rev. A} \textbf{\bibinfo{volume}{71}},
  \bibinfo{pages}{053605} (\bibinfo{year}{2005}).

\bibitem[{\citenamefont{Berg et~al.}(2012)\citenamefont{Berg, Rudner, and
  Kivelson}}]{Berg2012}
\bibinfo{author}{\bibfnamefont{E.}~\bibnamefont{Berg}},
  \bibinfo{author}{\bibfnamefont{M.~S.} \bibnamefont{Rudner}},
  \bibnamefont{and} \bibinfo{author}{\bibfnamefont{S.~A.}
  \bibnamefont{Kivelson}}, \bibinfo{journal}{Phys. Rev. B}
  \textbf{\bibinfo{volume}{85}}, \bibinfo{pages}{035116}
  (\bibinfo{year}{2012}).

\bibitem[{\citenamefont{Mitra et~al.}()\citenamefont{Mitra, Williams, and
  de~Melo}}]{Mitra}
\bibinfo{author}{\bibfnamefont{K.}~\bibnamefont{Mitra}},
  \bibinfo{author}{\bibfnamefont{C.~J.} \bibnamefont{Williams}},
  \bibnamefont{and} \bibinfo{author}{\bibfnamefont{C.~A. R.~S{\' a}}
  \bibnamefont{de~Melo}}, \bibinfo{note}{arXiv:0903.4655 [cond-mat.other]}.

\bibitem[{\citenamefont{Nelson and Halperin}(1979)}]{Nelson1979}
\bibinfo{author}{\bibfnamefont{D.~R.} \bibnamefont{Nelson}} \bibnamefont{and}
  \bibinfo{author}{\bibfnamefont{B.~I.} \bibnamefont{Halperin}},
  \bibinfo{journal}{Phys. Rev. B} \textbf{\bibinfo{volume}{19}},
  \bibinfo{pages}{2457} (\bibinfo{year}{1979}).

\end{thebibliography}

\end{document}